\documentclass[a4paper,aps,twocolumn,superscriptaddress,11pt,accepted=2017-07-25]{quantumarticle}

\usepackage[latin1]{inputenc}
\usepackage{amsmath, amssymb, amsfonts, mathrsfs}
\usepackage{graphicx}

\usepackage{tikz}
\usepackage{lipsum}

\graphicspath{{images/}}
\usepackage{xcolor}
\usepackage{exscale}
\usepackage{graphicx}
\usepackage{amsmath}
\usepackage{latexsym}
\usepackage{amsfonts}
\usepackage{amssymb}
\usepackage{bbm}

\def\pmx{\begin{pmatrix}}
\def\emx{\end{pmatrix}}

\newcommand{\ket}[1]{|#1\rangle}
\newcommand{\bra}[1]{ \langle #1 \,  |}
\newcommand{\braket}[1]{ \langle #1 \, \rangle}

\newcommand{\roof}[1]{ \left\lceil #1 \right\rceil}
\newcommand{\floors}[1]{ \left\lfloor  #1 \right\rfloor}

\begin{document} 

\title{Quantifying high dimensional entanglement with two mutually unbiased bases}

\author{Paul Erker}
\affiliation{Universitat Autonoma de Barcelona, 08193 Bellaterra, Barcelona, Spain}
\affiliation{Faculty of Informatics, Universit\`{a} della Svizzera italiana, Via G. Buffi 13, 6900 Lugano, Switzerland}
\affiliation{Facolt\`{a} indipendente di Gandria, Lunga scala, 6978 Gandria, Switzerland}
\author{Mario Krenn}
\affiliation{Vienna Center for Quantum Science and Technology (VCQ), Faculty of Physics, University of Vienna, Boltzmanngasse 5, A-1090 Vienna, Austria.}
\affiliation{Institute for Quantum Optics and Quantum Information (IQOQI), Austrian Academy of Sciences, Boltzmanngasse 3, A-1090 Vienna, Austria.}
\orcid{0000-0003-1620-9207}
\author{Marcus Huber}
\affiliation{Group of Applied Physics, University of Geneva, 1211 Geneva 4, Switzerland}
\affiliation{Universitat Autonoma de Barcelona, 08193 Bellaterra, Barcelona, Spain}
\affiliation{ICFO-Institut de Ciencies Fotoniques, 08860 Castelldefels, Barcelona, Spain}
\affiliation{Institute for Quantum Optics and Quantum Information (IQOQI), Austrian Academy of Sciences, Boltzmanngasse 3, A-1090 Vienna, Austria.}
\orcid{0000-0003-1985-4623}
\begin{abstract}
We derive a framework for quantifying entanglement in multipartite and high dimensional systems using only correlations in two unbiased bases. We furthermore develop such bounds in cases where the second basis is not characterized beyond being unbiased, thus enabling entanglement quantification with minimal assumptions. Furthermore, we show that it is feasible to experimentally implement our method with readily available equipment and even conservative estimates of physical parameters. 
\end{abstract}
\maketitle

Entanglement has long been recognized as the key concept that takes quantum communication beyond the classically possible. It finds applications in secure key distribution \cite{ekert91}, super-dense coding \cite{sdc} and improves communication capacities in a general sense \cite{eac}.\\
So far, most photonic implementations rely on two-dimensional degrees of freedom, limiting the capacity of each exchanged photon to one bit of information. Recent progress in understanding high dimensional degrees of freedom of photons, however, revealed the high capacity of entanglement in photon pairs naturally emerging from down-conversion processes. Prominent examples are path entangled photons in waveguides \cite{path1,path2,path3,path4}, access to the entangled angular momentum of photon pairs \cite{am1,amm,amm2,am2,agnew,besselEnt,am3,am4} and energy-time binned photons \cite{tb1,tb2,tb3}. Such high dimensional entanglement endows each photon pair with more shared information, greatly improving the efficiency of known protocols and enabling secure quantum communication at noise levels that would be prohibitive for qubit systems \cite{hd1,hd2,hd3,hd4,hd5}.\\
A lot of attention has been devoted to proving entanglement for such high dimensional systems \cite{science}, revealing the underlying dimensionality of entanglement \cite{path3,am2,am4} or generally characterizing the potential for accommodating many dimensions \cite{shannon0,shannon1,shannon2,shannon3,shannon4}. These initial investigations reveal the clear potential of the underlying systems, but in order to properly quantify the advantage provided one needs to actually quantify the number of entangled bits (e-bits, which is the number of two-dimensional Bell states that are necessary to produce the state) shared by the photons. The dimensionality of entanglement (i.e. the Schmidt rank) is given by the rank of the reduced density matrix and denotes the minimal dimension that is needed to reproduce the correlations of the state, whereas the e-bits are given by the entropy of that reduction, giving a clear operational meaning to the amount of information encoded in the entanglement. For example $\ket{\psi}=\frac{1}{\sqrt{2+k\epsilon^2}}\left( \ket{0,0}+ \ket{1,1} + \epsilon\sum_{i=1}^k\ket{i,i}\right)$ with $\epsilon$ being very small illustrates this difference. While this state is $k+2$-dimensional entangled (i.e. Schmidt rank $k+2$), it takes only little more than a single Bell state to create it (which shows up in its entanglement entropy being close to one as long as $k\epsilon^2\ll1$). On the other hand, this state's Schmidt number is very fragile with respect to noise, so experimentally certified Schmidt-numbers need more entangled bits to be robust. Calculating entangled bits for general states, however, is a notoriously difficult task \cite{review} and even at full access to a reconstructed density matrix there is no known method for computing this number in an efficient way (the best known algorithm just for deciding whether it is nonzero is exponential in the system's dimension \cite{DPSnew}). While entanglement witnesses in general only provide an answer to the question whether a given state exhibits entanglement \cite{db}, their actual value can also be used to quantify the amount of entanglement \cite{wq0,wq0b,wq1,wq2,wq3,wq4,review}. Unfortunately, generic entanglement witnesses in this context require a number of local measurement bases that scales with the system size, thus increasing the complexity rapidly with a growing number of degrees of freedom. Using mutually unbiased bases on the other hand provides a means of revealing entanglement with just two local measurements \cite{spengi}, a fact that has also been exploited for high dimensional experiments \cite{trixi,belleza,belleza2}.\\

In this work we combine the advantage of both approaches and quantify high dimensional entanglement purely from correlations in two mutually unbiased bases (MUBs), and show that the resulting quantification can readily be implemented experimentally with modern cameras. In fact, for sufficiently pure states the entire high dimensional entanglement can be certified with this minimal access. Since with only two measurement settings the word mutually seems superfluous we will sometimes only refer to unbiased bases. A natural candidate for two such bases are discretised position and momentum correlation. They are known to be readily accessible at a high quality and have thus been used to ascertain entanglement before \cite{posmom1,posmom2,posmom3}. Therefore, to showcase our theorem we derive how modern cameras and lenses (which perform a Fourier transformation in the far field/focus) can be harnessed to quantify the spatial correlations in down-conversion photons.  After treating the bipartite case we move on to multipartite systems and show how to quantify multipartite entanglement using two local unbiased measurement settings.\\
Before we do, let us introduce the relevant concepts. Sets of basis vectors $\{|v_i^k\rangle\}$ are called (mutually) unbiased, if they are both orthonormal $\langle v_i^k|v_j^k\rangle=\delta_{ij}$ and their overlaps are unbiased $|\langle v_i^k|v_j^{k'}\rangle|^2=\frac{1}{d}$. MUBs can be used for an efficient tomography \cite{woot}, cryptography protocols \cite{cerf} and in prime power dimensions there exist exactly $d+1$ such bases (see \cite{durt} for a review and further applications). It is still an unsolved problem how many MUBs exist in general (the smallest example being $d=6$). Our intention however is to make use of only two such bases, which always exist for any dimension $d$.\\
To quantify entanglement we use entanglement of formation (EOF) \cite{EOF}, which for pure states quantifies the asymptotic conversion rate between maximally entangled states and the quantum state under investigation. For mixed states its regularised asymptotic version quantifies precisely this entanglement cost as a rate of "`target states per Bell state"'. In other words, given $N$ copies of qubit Bell states, how many copies $k$ of the target state can we deterministically create using only local operations and classical communication (LOCC)? This asymptotic conversion rate $\frac{N}{k}$ is then found in the limit $N \to \infty$. For pure states it corresponds to the entropy of entanglement $E_{oF}(|\psi_{AB}\rangle):=S(\text{Tr}_{A/B}(|\psi_{AB}\rangle\langle\psi_{AB}|))$, where $S(\rho)=-\text{Tr}(\rho\log(\rho))$. For general quantum states Entanglement of Formation can be evaluated via a convex roof construction as the minimal average entanglement across all possible decompositions $E_{oF}(\rho)=\inf_{\mathcal{D}(\rho)}\sum_ip_iE_{oF}(|\psi_i\rangle)$. Even if the whole state $\rho$ is known exactly, it is a hard problem even to decide whether the measure is nonzero \cite{gurvits}, but we now want to find a lower bound from only two measurement outcomes.\\
The central figure of merit will be similar to the ones developed in \cite{spengi}, such that we can easily apply our methods to already existing experimental data (e.g. from \cite{belleza,trixi}). For bipartite systems the existing method makes use of the sum over all diagonal correlations in $m$ different MUBs, i.e.
$C_{m}(\rho)=\sum_{k=1}^{m}\sum_{i=0}^{d-1}\langle v_i^k (v_i^k)^*|\rho|v_i^k (v_i^k)^*\rangle\,.$
The simplest example are polarisation encoded states, for which $m=3$ MUBs can be used to compute $C_3=\frac{\braket{H,H}+\braket{V,V}+\braket{D,D}+\braket{A,A}+\braket{L,L}+\braket{R,R}}{\braket{H,H}+\braket{H,V}+\braket{V,H}+\braket{V,V}}$, where $\braket{X,Y}$ denotes coincidence counts in $X$ and $Y$ for the first and second photon, respectively and $H,V,D,A,L,R$ denote the six different polarisation states. This quantity is known to be bounded for separable states by $C_m(\rho_{sep})\leq 1+\frac{m-1}{d}$, whereas maximally entangled states can reach a value of $m$ \cite{spengi}. We are interested in the case $m=2$ as it's the minimal number of MUBs that can be used to verify entanglement, and it is experimentally the simplest case to access. Importantly, for large classes of states (and all pure entangled states), $C_2$ is already sufficient to detect entanglement. But how much can we say about the ability to give quantitative bounds on entanglement measures, such as EOF using only two bases?

A motivating example is provided in an idealised two qubit case: The eigenstates of the Pauli matrices $\sigma_i$ form MUBs. If we now measured correlations in such as e.g. $\langle \sigma_x\otimes\sigma_x\rangle$ and $\langle \sigma_y\otimes\sigma_y\rangle$ and found both of these values to be (close to) $-1$, positivity of the density matrix implies that $\langle \sigma_z\otimes\sigma_z\rangle\approx-1$. So despite only having measured two out of the three defining correlations of a qubit Bell state, we can infer that the state is indeed close to a Bell state and thus it's entanglement is close to $1$. Indeed, such an example is given in Ref.~\cite{toth} along with a semi-definite programming characterisation (SDP) to evaluate the convex roof extended linear entropy, even if the values are not close to $-1$. This further motivates two questions that naturally follow from this rather idealised setting: 
\begin{itemize}
	\item Can we make these considerations analytical, noise robust and quantitative?
	\item Are two measurements still sufficient for any dimension?
\end{itemize}
We affirmatively answer both questions by introducing the following quantity
\begin{widetext}
\begin{equation}\label{equation2}
\begin{split}
B(\rho)= N\Big[ d \left(\sum_{i=0}^{d-1}\langle v_i^2 (v_i^2)^*|\rho|v_i^2 (v_i^2)^*\rangle \right)-1-\sum_{\substack{m\neq n, m \neq l \\ l \neq o, n \neq o}} \sqrt{\langle v_m^1v_{n}^1|\rho|v_m^1v_{n}^1\rangle\langle v_{l}^1v_{o}^1|\rho|v_{l}^1v_{o}^1\rangle}\\-\sum_{i \neq j}\sqrt{\langle v_i^1 v_j^1|\rho|v_i^1 v_j^1\rangle\langle v_j^1 v_i^1|\rho|v_j^1 v_i^1\rangle\,}\Big]
\end{split}
\end{equation}
\end{widetext}
where we have used $N=\sqrt{\frac{2}{d(d-1)}}$, and the additional terms can be recorded alongside with the original measurements (without the need of adjusting the local measurement settings).
Our main result is the fact that $B(\rho)$ is indeed a direct lower bound to the generalised concurrence and thus a lower bound on the Entanglement of Formation can be easily computed whenever $B(\rho)\geq0$, via 
\begin{align}
E_{oF}(\rho)\geq -\log(1-\frac{B(\rho)^2}{2})\,.
\label{EoF}
\end{align}
The details of this derivation can be found in the appendix (\ref{sec:der}). We want to stress that this result holds for any choice of the two MUBs. One immediate consequence is that with just two global measurement settings we can certify that the maximally entangled state $|\phi^+\rangle=\frac{1}{\sqrt{d}}(\sum_{i=0}^{d-1}|ii\rangle)$ has an entanglement of formation of $E_{oF}=\log(d)$, i.e. all the entanglement can be quantified exactly through $B(|\phi^+\rangle\langle\phi^+|)=\sqrt{2 (1-\frac{1}{d})}$.\\
While at first glance it seems surprising that an entanglement of $\log(d)$ can be certified using only two measurement settings, there is of course a trade-off. The extra terms required make the bounds more sensitive to noise. So in order to gauge the practical usefulness it will be essential to study the performance of the bound in experimentally feasible settings with all sources of noise taken into account. But before we continue our discussion on experimental applicability and an analysis on the noise robustness we now give a brief excursion into multipartite variants of this theorem. Quantifying multipartite entanglement is a notoriously hard task, as there is no unique "currency", from which every state can be created via LOCC. Recent progress has been made using maximally entangled sets \cite{MES1,MES2}, however it remains an open problem to evaluate measures based on that concept on many high dimensional parties. We will make use of a more crude classification, capturing necessary resources, but leaving open the sufficiency entirely. Here we can make use of a multipartite generalization of entanglement of formation proposed in \cite{wq2,maju,mamaju}. The operational interpretation is simply the minimal necessary average entanglement across every cut for creating this state via LOCC. It does not reveal a deeper structure, but is nonzero for every genuinely multipartite entangled state and zero if there exists a decomposition into at least biseparable states. Formally it can be defined as
\begin{align}
E_{GME}:=\inf_{\mathcal{D}(\rho)}\sum_ip_i \min_{A_i}S(\text{Tr}_{A_i}(|\psi_i\rangle\langle\psi_i|))\,.
\end{align}
\begin{figure*}[ht!]
\includegraphics[width=\textwidth]{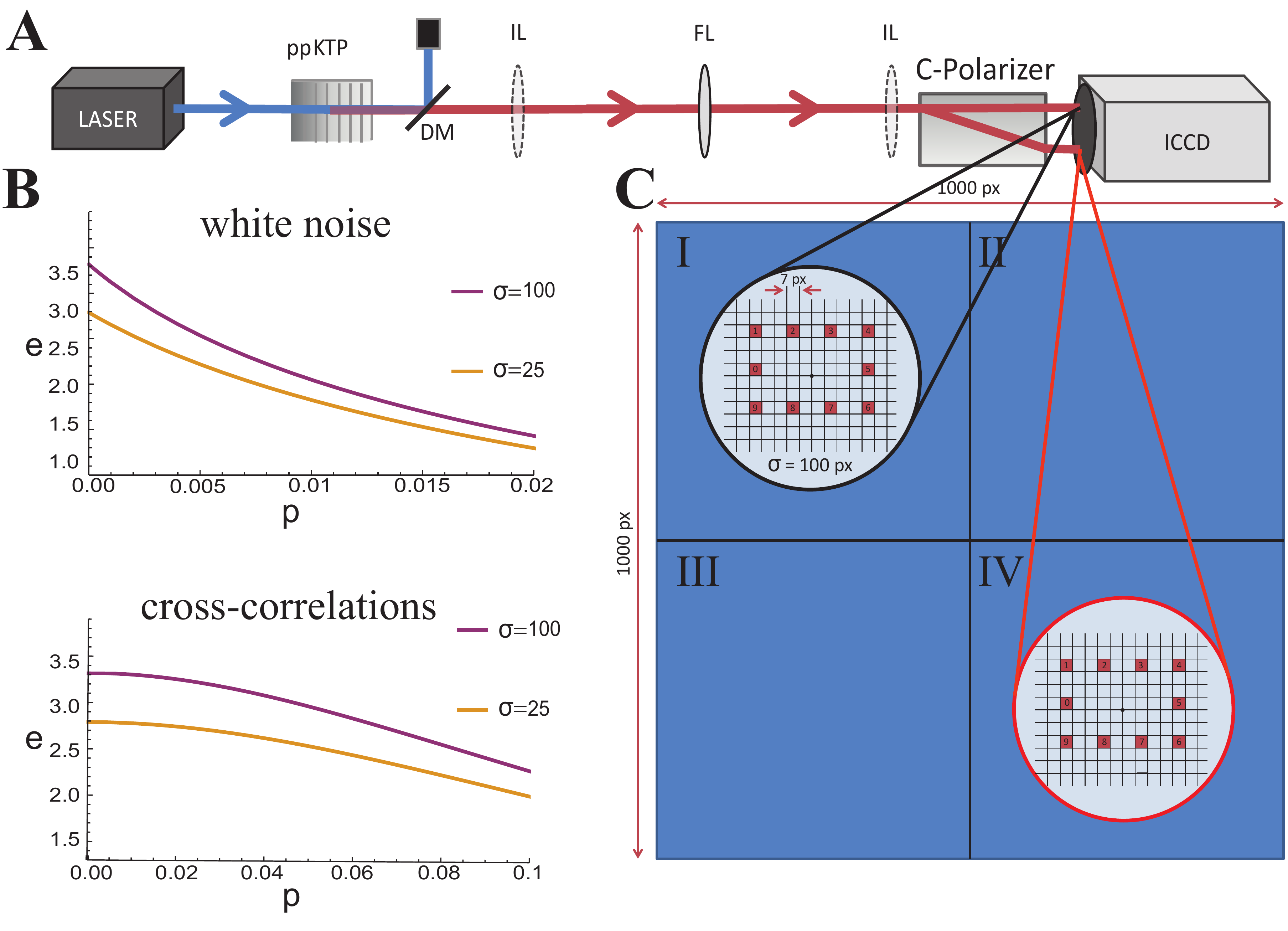}\\
\caption{\textbf{A}: The proposed experimental setup. A laser produces pairs of spatially entangled photons with wavelength of 800nm in a SPDC process in the nonlinear ppKTP crystal. After the pump beam is removed by a dicroic mirror, the photon-pair is deterministically separated with a calcit polarizer. The two parts of the beam then go to different positions at a camera. Depending on the lens configuration, the photons are detected in the position-basis (if the two imaging lenses (IL) are in the beam path) or in the momentum-basis (if only the Fourier lens (FL) is used). \textbf{B}: Here we showcase two different effects that predominantly reduce the certifiable e-bits, white noise and cross-correlations between different pixels. The purple and orange graphs show different weighting of the individual terms. Purple depicts a beam with $\sigma=100$ pixels, and orange $\sigma=25$ pixels. $p$ stands for the probability of white noise or cross-correlations, respectively. \textbf{C}: The considered areas of pixels at the screen of the ICCD camera as considered in the calculations. Due to the significant impact of white noise, it is optimal to use only small parts of the camera. }  
\label{fig:AllPix}
\end{figure*}

While, again, two mutually unbiased bases can suffice for some states to quantify all multipartite entanglement exactly, the multitude of potential correlation structures giving rise to Genuine Multipartite Entanglement (GME) makes the construction of one universal criterion impossible. While the criteria themselves constitute genuine lower bounds and need no assumption about the underlying state, their capability of revealing multipartite entanglement will be tailored to specific classes of states. In fact the same is true for bipartite systems, however, with a change of local basis every pure state can be written as $|\psi\rangle=\sum_{i=0}^{r_S}\lambda_i|v_iv_i^*\rangle$. This Schmidt decomposition is not possible for multipartite states, so our criteria will have to be adapted to the target state in question. In the appendix we therefore describe a generic recipe how this can be accomplished and illustrate it with known states of a specific dimensionality structure \cite{linalg}, that can be readily produced in optical setups \cite{mehul}.

In order to explore the feasibility of our criterion we come back to bipartite systems and the issue of noise resistance. As mentioned before, the generality of bounding entanglement through any arbitrary pair of MUBs, without any information on the relative phases comes at the price of high noise sensitivity. The amount of noise that can be tolerated, of course, depends on the nature of the noise itself. To start, let us analyse two paradigmatic noise models and how they theoretically impact the bound. The first is dephasing noise, i.e. a noise model where the intended (maximally entangled) target state is mixed with its dephased version:
\begin{equation}
\rho_{dp}(p)=p|\phi^+\rangle\langle\phi^+|+\frac{(1-p)}{d}(\sum_{i=0}^{d-1}|ii\rangle\langle ii|)\,.
\end{equation}
In this case $B(\rho_{dp})=N(d (p+\frac{(1-p)}{d^2}-1)$, which implies entanglement detection up to a noise threshold of $p_{crit}=\frac{1}{d+1}$, which actually increases noise resistance with the dimension $d$. The other extreme case is of course white noise, which provides the toughest challenge for entanglement detection.
\begin{equation}
\rho_{wn}(p)=p|\phi^+\rangle\langle\phi^+|+\frac{(1-p)}{d^2}\mathbbm{1}\,.
\end{equation}
Here $B(\rho_{wn})=N(d (p+\frac{(1-p)}{d^2}-1-\frac{d-2}{d}(d-1)^2(1-p))$, which leads to a decreasing noise resistance of $p_{crit}=\frac{d^2-3d+2}{d^2-2d+2}$ which approaches $1$ for $d\rightarrow\infty$. Both of these noise models are somewhat unrealistic, however, and the true practicality can only be ascertained through an experimental proposal.

We therefore consider a scenario where a pair of photons illuminate a single-photon sensitive camera. With simple optics, we can access two mutual unbiased bases (the position-basis and the momentum-basis), which we will use to estimate the strength and noise-dependence of the criterion. This is the ideal testbed, as it also plays on the strength of not needing to know the exact phase relation between the mutually unbiased bases.

The photon pairs are created in a nonlinear spontaneous parametric down-conversion (SPDC) crystal \cite{Walborn2}, which is the working horse in quantum optical experiments. The two photons, which are deterministically separated (for instance when they have opposite polarization in a type-II SPDC, or different wavelengths), illuminate two different regions of an ICCD (intensified charge-coupled device) camera. ICCDs have high detection efficiencies (20\% for 800nm, up to 50\% for green) and have a very low dark count rate, which makes it possible to identify single photons and photon pairs. (This is in contrast to electron multiplying charge-coupled device (EMCCD) cameras, which have high efficiencies but high dark count rates. There, quantum correlations can be inferred via averaging techniques \cite{Luntz,Edgar}). In the first measurement, the crystal is imaged onto the camera, which results in strong position correlations. The second measurement is performed using a lens to go to the Fourier-plane of the crystal, which results in strong momentum correlations - thus in strong position anti-correlations. The strength of spatial correlations can be quantified with the Federov-Ratio $F$, which is the ratio between the marginal and conditional probability \cite{Walborn1, Walborn2}. Experimentally, one can reach very large values - realistically $F$ could be $25$ \cite{Walborn2}, which is the value we calculate with .

To estimate the detectable number of e-bits from (\ref{EoF}), one has to consider several effects that reduce the extractable e-bits: First, dark counts are roughly evenly distributed over the camera pixels, resulting in white noise. Secondly, cross-correlations between different states further reduce the certifiable entanglement. And thirdly, non-equal weighting between different modes are influencing the final result as well. The interplay of these effects can be seen in fig.(\ref{fig:AllPix}B). 

Different camera configurations will lead to a differently sized influence of these effects. Larger regions grouped together reduce cross-talk, but will increase dark counts per region. However, with careful considerations of the influencing parameters, fig.(\ref{fig:AllPix}B), one can nonetheless find configurations that can certify more than $3$ e-bits, with only two MUBs. These values are reached entirely without phase characterization of the MUBs. If we restrict ourselves to areas of $7\times7$ pixels, we find that the optimal number of areas to be considered is $10$, where we find a noise-level of $0.6\%$. If we use more areas, the white noise probability increases (as more pixels are considered). With this information, we find a good configuration which reduces as much as possible the negative influence of the other two effects. Specifically, we introduce large empty areas between the considered pixels in order to reduce cross-talk. While this will reduce the number of photons detected per unit of time proportional to the area not counted, it will strongly increases the certified e-bits. This is because of the assumed large single-photon beam, where the probability to find a photon in any of the different areas is very similar. In the end, we find that this configuration leads to $2.4$ e-bits, mainly reduced due to white noise. If we decrease the number of pixels further to $3\times3$ areas, we find $3.05$ e-bits. For the reader's convenience we present a detailed step-by-step calculation of the $d=3$ case (i.e. $3$ areas) in the appendix.

Except for the careful design of the optimal areas for correlations (which can be done after the data is collected), another experimental challenge is the stabilization of the setup for longer times. The ICCD camera has a limit of roughly $10$ images per second. As one wants to reduce multi-photon events (as they would increase the white-noise fraction), the photon production rate has to be set at a very low level to roughly $1$ Hz. Only one pair in $100$ arrives at the considered areas - together with the detection efficiency of both photons, it results in roughly $10^{-4}$ Hz of detected photon pairs. In order to collect a suitable amount of correlated photons, one could expect the measurement time to be $100-200$ hours. This caveat could be addressed in practice by using compressed sensing approaches, as e.g. in \cite{posmom3}. An alternative way to improve the extractable e-bits is the shift from the standard 800nm photon pairs (for which the camera has an efficiency of 20\%) to green entangled photons \cite{greenent} which can be detected with an efficiency of 50\%. Also the useage of novel sCMOS cameras \cite{cmos1, cmos2}, which are faster and have lower noise compared to the ICCD camera considered here, will improve the measurement time and extractable ebits. This concludes the analysis of the feasibility of using only two MUBs to experimentally quantify entanglement.

While in principle any amount of bipartite entanglement can be certified for pure states, realistic noise assessment shows that with this limited number of local observables and current technology we can still certify more than three times the entanglement of a perfect qubit entangled pair. We hope that this number can still be increased through rapidly developing camera technology or possibly even through a more clever detection design. While this method serves as a technique to experimentally prove the potential of down conversion sources for high dimensional quantum communication, there are still many open questions. Especially quantum key distribution protocols display a strong trade-off between security and implementability. In particular fully device independent quantum key distribution requires (almost) loophole free Bell inequality violation, a feat that is hard to achieve at sensible key rates. Prepare and measure schemes can be easily attacked if the source or measurement devices are hacked. It may be possible to use entanglement quantifiers, such as the one presented here, to certify security of the source for high dimensional QKD, interpolating between prepare and measure and fully device independent schemes. Another path to pursue would of course be the inclusion of further measurements to strengthen the resistance to noise, which in our experimental proposal would also require a precise phase control in the Fourier-plane pixels. Our method can also be implemented for other quasi-continuous variable entangled systems. In particular it might be particular useful for wavelength-entangled systems, where access to two MUBs (wavelength and its complementary, time), is possible - but higher numbers of MUBs are not easily accessible. There our method might be able to certify the great potential of wavelength-entangled quantum systems\cite{timeenergy1, timeenergy2, timeenergy3}.

\emph{Acknowledgements} We are grateful to Robert Boyd, Anton Zeilinger, Ebrahim Karimi, Robert Fickler and Mehul Malik for discussions that started and shaped this project. We thank Claude Kl\"ockl for productive discussions at LIQUID and Robert Fickler, Armin Hochrainer and Nicolai Friis for helpful feedback. MH and PE were supported by the European Commission (STREP ``RAQUEL''), by the Spanish MINECO, projects FIS2008-01236 and FIS2013-40627-P, with the support of FEDER funds, and by the Generalitat de Catalunya CIRIT, project 2014-SGR-966. PE furthermore acknowledges funding from the Swiss National Science Foundation (SNF) and the National Centres of Competence in Research "Quantum Science and Technology" (QSIT). MH furthermore acknowledges funding from the Juan de la Cierva fellowship (JCI 2012-14155), the Swiss National Science Foundation (AMBIZIONE PZ00P2$\_$161351) and the Austrian Science Fund (FWF) through the START project Y879-N27. MK acknowlegdes support from the European Research Council (SIQS Grant No. 600645 EU-FP7-ICT) and the Austrian Science Fund (FWF) with SFB F40 (FOQUS).

%
\onecolumngrid

\section{Analysing the noise robustness}\label{noisy}
First we analyse the noise robustness of the criterion presented in the main text in a realistic experimental scheme. Because of three different experimental effects, the detected state differs from a maximally entangled high-dimensional pure state: Unequal weighting due to gaussian shape of the photons' spatial distribution, cross correlations due to non-perfect correlations, and white noise due to accidental dark counts of the camera.

\subsection{Unequal weighting}
The unequal weighting of the different pixel-areas is mainly due to the Gaussian character and the discretisation of the grid. The state can can be written as

\begin{align}      
\left|\psi\right\rangle=N_1 \Big(\sum_{x=0}^{x_{max}} w_x \overline{\left| x,x\right\rangle} \Big)   
\label{psiWeight}
\end{align}
where $w_x$ can be found numerically and $\overline{\left| x,x\right\rangle}$ is explained in the next section. The weighting could be flattened by adjusting the size of the discretisation. However, as we will show, the white noise contributions is significant for large number of pixel-areas.

\subsection{Cross correlations}
Due to physical constraints in the production of photon pairs (such as the size of the pump beam or the length of the crystal), the spatial correlations can not be infinitely strong. It results in cross-correlations between pixel areas that reduce the entanglement.

We consider only first-order contributions from cross correlations (which means, only direct neighboring pixel areas contribute). The cross-correlations depend on the geometry of the considered pixels. In Fig. 1C in the main text, one can see that pixel all pixels have two direct neighbors. In general, one finds:
\begin{align}  
  \overline{\left| x,x\right\rangle}=N_2 \Big( c_x \left| x,x\right\rangle +  \sum_{y\in NB(x)} c_{x,y} \big( \left| x,y\right\rangle + \left| y,x\right\rangle \big) \Big)
\end{align}


where $c_x$ and $c_{x,y}$ are found numerically, and $NB(x)$ is the set of neighbors of $x$. One could significantly reduce the off-correlations by introducing unobserved pixel-lines between the $N\times N$-areas, as shown in Fig. 1C in the main text.

\subsection{White Noise}
The dominating source of white noise comes from photon loss and from dark counts introduced in the measurement of the ICCD camera. As the loss of the photons and the dark counts are evenly distributed over the camera area, we can model them as following:
\begin{align}
\rho=p|\psi\rangle\langle\psi|+\frac{1-p}{d^2}\mathbbm{1}_{d^2}\,,
\end{align}
where $|\psi\rangle$ is defined in eq. (\ref{psiWeight}). 

The white noise comes from dark counts, and from multi-pair detection from the crystal. This is not negligible, because the integration time is 0.1sec. For a given setting of dark count-rate, detector-efficiency and used area of the ICCD, there is an optimal value of the photon-pair rate P (which is very low, roughly $10^{-2}$ per seconds).

We only consider camera images with exactly two photons, one in region A and one in region B. All other events are rejected. Such two photon events can happen in the following ways:
\begin{enumerate}
\item Two dark counts; one in Region 1 (where photon A usually appears), one in Region 2
\item One dark count, one real count
\item Two real counts
\end{enumerate}

Now a simple calculation shows:
\begin{align}      
  \textnormal{all counts}&= \bar{P} \left(D_1 D_2 \right) +\\ \nonumber 
  &+ P^1\left(D_1 D_2 \bar{\epsilon_1} \bar{\epsilon_2} + D_1 \epsilon_2 \bar{D_2} \bar{\epsilon_1} + \epsilon_1 D_2 \bar{D_1} \bar{\epsilon_2} + \epsilon_1 \epsilon_2 \bar{D_1} \bar{D_2} \right) +\\\nonumber
  &+ P^2\left(D_1 D_2 \bar{\epsilon_1}^2 \bar{\epsilon_2}^2 + 2 D_1 \epsilon_2 \bar{D_2} \bar{\epsilon_1}^2 \bar{\epsilon_2} + 2 \epsilon_1 D_2 \bar{D_1} \bar{\epsilon_2}^2 \bar{\epsilon_1} + 4 \epsilon_1 \epsilon_2 \bar{D_1} \bar{D_2} \bar{\epsilon_1} \bar{\epsilon_2} \right) +\\ \nonumber
  &+ ...\\ \nonumber
  &= \sum_{n=0}^{\infty} P^n \left(D_1 D_2 \bar{\epsilon_1}^n \bar{\epsilon_2}^n + n D_1 \epsilon_2 \bar{D_2} \bar{\epsilon_1}^n \bar{\epsilon_2}^{n-1} + n \epsilon_1 D_2 \bar{D_1} \bar{\epsilon_2}^n \bar{\epsilon_1}^{n-1} + n^2 \epsilon_1 \epsilon_2 \bar{D_1} \bar{D_2} \bar{\epsilon_1}^{n-1} \bar{\epsilon_2}^{n-1} \right)
\end{align}
where the last sum has a simple closed form and $\bar{P}$ is the probability that no photon-pair is created. The counts we are interested in are two real counts, without dark counts:

\begin{align}      
  \textnormal{good counts}&= P^1\left(\epsilon_1 \epsilon_2 \bar{D_1} \bar{D_2} \right)
\end{align}

\begin{align}      
  p_{good}= \frac{\textnormal{good counts}}{\textnormal{all counts}}
\end{align}

We use $D_1=D_2$ (dark counts in the two areas are the same) and $\epsilon_1=\epsilon_2$ (the efficiency in the two areas are the same). The white-noise probability is $p_{white noise}=1-p_{good}$.

\section{Detailed derivation for bipartite systems}\label{sec:der}
We now want to present a detailed derivation of the fact that two MUB measurements can quantify entanglement, even without any knowledge about the detailed structure of the second MUB. All one needs is to make sure that the outcomes are indeed unbiased. We start the derivation by reminding the reader of the lower bounds for the concurrence (i.e. the square root of the linear entropy) developed in \cite{maju,mamaju} as
\begin{align}
I:=\sqrt{\frac{2}{d(d-1)}}\left( \sum_{m\neq n}\underbrace{|\langle mm|\rho|nn\rangle|}_{I_1}-\underbrace{\sqrt{\langle mn|\rho|mn\rangle\langle nm|\rho|nm\rangle}}_{I_2} \right) \leq\inf_{\mathcal{D}(\rho)}\sum_ip_i\sqrt{(2(1-\text{Tr}({\rho_A^i}^2))}\,.
\end{align}
To get a bound on entanglement of formation we can use the relation between the linear entropy and the Renyi 2-entropy $S_2(\rho)=-\log(\text{Tr}(\rho^2))$ and the fact that the family of Renyi entropies is monotonically decreasing in $\alpha$, i.e. that for $S_\alpha:=\frac{1}{1-\alpha}\text{Tr}(\rho^\alpha)$ it holds that $S_\alpha\geq S_\beta\,\forall \alpha\leq\beta$. This directly implies that entanglement of formation, defined as
\begin{align}
E_{oF}:=\inf_{\mathcal{D}(\rho)}\sum_ip_iS(\rho_A^i)
\end{align}
is directly lower bounded by 
\begin{align}
E_{oF}\geq -\log(1-\frac{I^2}{2})
\end{align}
So the task at hand is to experimentally estimate the bound $I$ with only two mutually unbiased measurements. The first thing to notice is that $I_2$ is directly accessible from correlations in the first basis, i.e. choosing $\{|v_1^i\rangle\}=\{|i\rangle\}$
\begin{align}
\langle mn|\rho|mn\rangle=\frac{N_{m,n}}{\sum_{i,j}N_{i,j}}\,,
\end{align}
where $N_{m,n}$ is just the total number of correlated clicks recorded between $m$ on Alice's side and $n$ on Bob's side. Since the term is strictly negative it is desirable to have these "wrong" correlations as suppressed as possible and thus define the basis in terms of the most correlated elements between Alice and Bob to be labeled by the same numbers $m$.\\
Now comes the more tricky part: Estimating the total number of coherences in the first term of $I$. First we notice that to be mutually unbiased to the computational basis every overlap $|\langle v^2_i|j\rangle|^2=\frac{1}{d}$, that means we can write the second basis as
\begin{align}
|v_k\rangle:=\frac{1}{\sqrt{d}}\sum_{m=0}^{d-1} e^{-i\phi_m^k}|m\rangle\,.
\end{align}
Now we can evaluate a specific sum of correlations in the mutually unbiased basis:
\begin{align}
\Sigma :=C_2 - C_1=\sum_{k}\langle v_k v_k^*|\rho|v_k v_k^*\rangle=\frac{\sum_k N_{k,k}}{\sum_{i,j}N_{i,j}}
\end{align}
we first notice that $|v_kv_k^*\rangle=\frac{1}{d}\sum_{m,n}e^{-i(\phi_m^{k}-\phi_{n}^{k})}|mn\rangle$, such that
\begin{align}
\Sigma=\frac{1}{d^2}\sum_{k}\sum_{m,n,l,o}e^{i(\phi_m^k-\phi_{n}^k+\phi_{o}^k-\phi_{l}^k)}\langle m n|\rho| lo\rangle
\end{align}
Now we can use that $\sum_{m=0}^{d-1} e^{i(\phi_m^{k}-\phi_m^{k'})}=0\,\forall \, k\neq k'$ due to the fact that $\langle v_k^2|v_{k'}^2\rangle=0\,\forall\,k\neq k'$. The sum $\Sigma$ can be split in three terms $\Sigma=\Sigma_1+\Sigma_2+\Sigma_3$:
\begin{itemize}
\item $m=l$ and $n=o$: This amounts to $\Sigma_1=\frac{1}{d}\underbrace{\sum_{m,n}\langle mn|\rho|mn\rangle}_{=\text{Tr}(\rho)=1}=\frac{1}{d}$
\item $m=n$ and $l=o$: This is exactly the desired sum, i.e. $\Sigma_2=\frac{1}{d}\sum_{m\neq l} \langle mm|\rho|ll\rangle$
\item For the term $m=l$ and $n \neq o$:Here we find a pre-factor of $\sum_{k=0}^{d-1}e^{i(\phi_{o}^k-\phi_{n}^k)}=0$. The same for $m=n$ and $l\neq o$, $n=o$ and $m\neq l$ and $l=o$ and $m\neq n$.
\item The remaining terms that do not vanish fulfill with $m\neq n$, $m\neq l$, $n\neq o$ and $l\neq o$. terms for which $\sum_ke^{i(\phi_m^k-\phi_{n}^k+\phi_{o}^k-\phi_{l}^k)}=c_{m,n,l,o}$ yield $\Sigma_3=\frac{1}{d^2}\sum_{\substack{m\neq n, m \neq l \\ l \neq o, n \neq o}} c_{m,n,l,o}\Re e[\langle mn|\rho|lo\rangle]$
\end{itemize}
Now we can make use of additional experimental data taken in the first basis and use the Cauchy-Schwarz-inequality to show
\begin{align}
d \Sigma_3-\sum_{\substack{m\neq n, m \neq l \\ l \neq o, n \neq o}} \sqrt{\langle mn|\rho|mn\rangle\langle lo|\rho|lo\rangle}\leq 0
\end{align}
Finally since $\Re e[z]\leq |z|$ we can now state the main result here being:
\begin{align}
I_1\geq d\left(\Sigma-\frac{1}{d}-\frac{1}{d} \sum_{\substack{m\neq n, m \neq l \\ l \neq o, n \neq o}} \sqrt{\langle mn|\rho|mn\rangle \langle lo | \rho | lo \rangle}\right) \,.
\end{align}
and thus
\begin{align}
I_1 - I_2 \geq d \left( \Sigma-\frac{1}{d}-\frac{1}{d}\sum_{\substack{m\neq n, m \neq l \\ l \neq o, n \neq o}}\sqrt{\langle mn|\rho|mn\rangle\langle lo|\rho|lo\rangle}\right)-\sum_{m\neq n}\sqrt{\langle mn|\rho|mn\rangle\langle nm|\rho|nm\rangle}\,,
\end{align}
 such that we finally arrive at
\begin{align}
E_{oF}\geq -\log(1-\frac{B(\rho)^2}{2})\,.
\end{align}

\section{The multipartite case}
Note that in contrast to the bipartite case where our results hold for any pair of MUBs in the following we choose a particular pair of MUBs to facilitate the derivation:
\begin{align}
|\tilde{i}_k \rangle := \sum_{m=0}^{d-1}  \omega^{\tilde{i} m } | m \rangle \,,
\end{align}
where $\omega:=e^{\frac{2\pi i}{d}}$. Using these we first introduce the following linear combination of diagonal density matrix elements:

\begin{align}
C_{n,d} := \sum_{\alpha} f_{\alpha} \langle \tilde{k}_{\alpha}|  \rho |\tilde{k}_{\alpha}\rangle
\end{align}
where $\alpha = {i_1, \dots, i_n}$ is a multi-index with $i_1, \dots, i_n \in \{0, \dots, d-1 \} $. Furthermore
\begin{align}
 f_{\alpha} := \begin{cases} 1 \ &\mbox{if } 0 \leq s_{\alpha} \leq \floors{\frac{d}{4}} \\
-1 \ &\mbox{if } \roof{\frac{d}{4}} \leq s_{\alpha} \leq \floors{\frac{3d}{4}} \\
1 \ &\mbox{if } \roof{\frac{3d}{4}} \leq s_{\alpha} \leq d-1\end{cases}
\end{align}
where
\begin{align}
s_{\alpha} := \sum_{j=1}^{n} i_{j} \mod{d} .
\end{align}

Now we have that
\begin{align}
C_{n,d} = - g   +\frac{1}{\xi} \sum_{\gamma} \Re e \langle k_{\alpha}|  \rho | k_{\beta} \rangle 
\end{align}
where $\gamma := \{ \alpha, \beta | i \in \alpha, k \in \beta: \langle i_j| k_j \rangle = 0 \ \forall \ 1 \leq j \leq n  \}$. Moreover if one defines $p_l$ as the number of combinations for which $s_{\alpha} = l $ then
\begin{align}
 \frac{1}{\xi} = \frac{1}{2 d^{n}} \sum_{l=0}^{d-1} p_l  | \Re e (\omega^l)|
\end{align}
and 
\begin{align}
 g = 1- \frac{2p_0 }{ d^{n}} .
\end{align}
This quantity $C_{n,d}$ is reminiscent of the figure of merit in the bipartite case (i.e. $\sum_{i=0}^{d-1}\langle v_i^2 (v_i^2)^*|\rho|v_i^2 (v_i^2)^*\rangle$) as it features off-diagonal elements potentially involved in multipartite entangled states. In the following we will use a similar strategy, relating these elements together with diagonal matrix elements to lower bounds on concurrences for multipartite systems \cite{wq2,wq3,wq4,maju,mamaju}. 

\begin{align}
 C_{GME} \geq \sqrt{\frac{2}{ d(d-1)} } ( \sum_{j=0}^{d-1}\sum_{i\neq j}|\langle i|^{\otimes n}  \rho | j \rangle^{\otimes n} | -  \sum_{\kappa} P_{ij}^{\kappa})
\end{align}

where

\begin{align}
 P_{ij}^{\kappa} = \sqrt{ \langle i|^{\otimes n} \langle j|^{\otimes n} \Pi_{\kappa} \rho^{\otimes 2} \Pi_{\kappa} | i \rangle^{\otimes n}| j \rangle^{\otimes n} } .
\end{align}

We can now use the linear combination above such that

\begin{align}
 C_{GME} \geq \sqrt{\frac{2}{ d(d-1)} } ( \xi C_{n,d} - \overline{C}_{n,d} -  \sum_{\kappa} P_{ij}^{\kappa} )
\end{align}
with

\begin{align}
 \overline{C}_{n,d} :=  \xi C_{n,d} -  \sum_{j=0}^{d-1}\sum_{i<j}|\langle i|^{\otimes n}  \rho | j \rangle^{\otimes n} | - g .
\end{align}
Making use of the Cauchy-Schwarz inequality, we get our main result, i.e. for $\gamma' := \gamma / \{ \alpha, \beta | i \in \alpha, k \in \beta: \langle i_j| k_j \rangle = 0 \ , \ i_j = i_h \ , \ k_j = k_h \ \forall \ 1 \leq j,h \leq n   \}$ it holds that
\begin{align}
 C_{GME} \geq \sqrt{\frac{2}{ d(d-1)} } \left[ \xi \left(C_{n,d} - \sum_{\gamma'} \sqrt{ \langle k_{\alpha}|  \rho | k_{\alpha} \rangle \langle k_{\beta}|  \rho | k_{\beta} \rangle} \right)     + g -  \sum_{\kappa} P_{ij}^{\kappa} \right].
\end{align}
As mentioned in the main text, this procedure will strongly depend on the off-diagonal elements that are revealed through the MUBs and the tailoring of the lower bounds themselves for specific target states. As a demonstration of feasibility we focus on the tripartite case, but stress that in principle for each target state one could try to develop a similar approach.
\subsection{Explicit tripartite examples}

Here we give explicit examples for the resulting inequalities for the case of $n=3$, i.e. a tripartite scenario. As the examples become quite cumbersome with increasing $d$ we will restrict to explicitly stating only the cases $d=2$ and $d=3$.\\

\subsubsection{Three Qubits}
We start out with the known bound  for the genuine multipartite concurrence \cite{maju,mamaju}
\begin{align}
C_{GME} \geq B_{GME}(\rho):=& 2 ( \Re e \langle 111|\rho|000\rangle \nonumber\\
&- \sqrt{\langle 001|\rho|001\rangle \langle 110|\rho|110\rangle} - \sqrt{\langle 010|\rho|010\rangle \langle 101|\rho|101\rangle}- \sqrt{\langle011 |\rho|011\rangle \langle 100|\rho|100\rangle} )
\end{align}
and choose the following linear combination of diagonal density matrix elements
\begin{align}
C_{3,2} &= \langle +++|\rho|+++\rangle + \langle +--|\rho|+--\rangle + \langle -+-|\rho|-+-\rangle \\ 
&+ \langle --+|\rho|--+\rangle - \langle ++- |\rho|++-\rangle  - \langle +-+|\rho|+-+\rangle \nonumber\\
&- \langle -++|\rho|-++\rangle - \langle ---|\rho|---\rangle \nonumber\\
&= 2 ( \Re e \langle 111|\rho|000\rangle + \Re e \langle 001|\rho|110\rangle +  \Re e \langle 010|\rho|101\rangle + \Re e \langle 100|\rho|011\rangle )
\end{align}
This leads to the following bound
\begin{align}
 C_{GME} \geq C_{3,2}  - 4  \left( \sqrt{\langle 001|\rho|001\rangle \langle 110|\rho|110\rangle} + \sqrt{\langle 010|\rho|010\rangle \langle 101|\rho|101\rangle}+ \sqrt{\langle011 |\rho|011\rangle \langle 100|\rho|100\rangle} \right)\,.
\end{align}
Just as in the bipartite case, this bound gives the correct value for a pure GHZ state (i.e. $B_{GME}(|GHZ\rangle\langle GHZ|)=C_{GME}(|GHZ\rangle\langle GHZ|)$, for both explicit examples. For $d=2$ and thus $|GHZ\rangle=\frac{1}{\sqrt{2}}(|000\rangle+|111\rangle)$ this results in a resistance to white noise (i.e. the critical value of $p$ such that $\rho_{noise}:=p|GHZ\rangle\langle GHZ|+\frac{1-p}{d^3}\mathbbm{1}$ is still detected to be genuinely multipartite entangled ) of $p_{crit}=\frac{3}{5}$.

\subsubsection{Three Qutrits}
In the tripartite qutrit case it was shown that \cite{maju,mamaju}
\begin{align}
C_{GME} &\geq \frac{2}{\sqrt{3}} ( \Re e \langle 000|\rho|111\rangle  + \Re e \langle 111|\rho|222\rangle + \Re e \langle 222|\rho|000\rangle  \nonumber\\ 
&- \sqrt{\langle 001|\rho|001\rangle\langle 110|\rho|110\rangle} - \sqrt{\langle 010|\rho|010\rangle\langle 101|\rho|101\rangle}- \sqrt{\langle 100|\rho|100\rangle\langle 011|\rho|011\rangle} \nonumber\\ 
&- \sqrt{\langle 112|\rho|112\rangle\langle 221|\rho|221\rangle} - \sqrt{\langle 121|\rho|121\rangle\langle 212|\rho|212\rangle}- \sqrt{\langle 122|\rho|122\rangle\langle 211|\rho|211\rangle} \nonumber\\ 
&- \sqrt{\langle 002|\rho|002\rangle\langle 220|\rho|220\rangle} - \sqrt{\langle 020|\rho|020\rangle\langle 202|\rho|202\rangle}- \sqrt{\langle022|\rho|022\rangle\langle 200|\rho|200\rangle} )
\end{align}
The linear combination we will make use of is defined as
\begin{align}
C_{3,3} &= \langle \tilde{0}\tilde{0}\tilde{0}|\rho|\tilde{0}\tilde{0}\tilde{0}\rangle + \langle \tilde{1}\tilde{1}\tilde{1}|\rho|\tilde{1}\tilde{1}\tilde{1}\rangle + \langle \tilde{2}\tilde{2}\tilde{2}|\rho|\tilde{2}\tilde{2}\tilde{2}\rangle \\ 
&+ \langle \tilde{0}\tilde{1}\tilde{2}|\rho|\tilde{0}\tilde{1}\tilde{2}\rangle + \langle \tilde{1}\tilde{2}\tilde{0} |\rho|\tilde{1}\tilde{2}\tilde{0}\rangle  + \langle \tilde{2}\tilde{0}\tilde{1}|\rho|\tilde{2}\tilde{0}\tilde{1}\rangle \nonumber\\
&+ \langle \tilde{1}\tilde{0}\tilde{2}|\rho| \tilde{1}\tilde{0}\tilde{2}\rangle + \langle \tilde{0}\tilde{2}\tilde{1}|\rho|\tilde{0}\tilde{2}\tilde{1}\rangle + \langle \tilde{2}\tilde{1}\tilde{0}|\rho|\tilde{2}\tilde{1}\tilde{0}\rangle\nonumber \\
&- \langle \tilde{0}\tilde{0}\tilde{1}|\rho| \tilde{0}\tilde{0}\tilde{1}\rangle - \langle \tilde{0}\tilde{1}\tilde{0}|\rho|\tilde{0}\tilde{1}\tilde{0}\rangle - \langle \tilde{1}\tilde{0}\tilde{0}|\rho|\tilde{1}\tilde{0}\tilde{0}\rangle\nonumber \\
&- \langle \tilde{2}\tilde{2}\tilde{0}|\rho| \tilde{2}\tilde{2}\tilde{0}\rangle - \langle \tilde{2}\tilde{0}\tilde{2}|\rho|\tilde{2}\tilde{0}\tilde{2}\rangle - \langle \tilde{0}\tilde{2}\tilde{2}|\rho|\tilde{0}\tilde{2}\tilde{2}\rangle\nonumber \\
&- \langle \tilde{1}\tilde{1}\tilde{2}|\rho| \tilde{1}\tilde{1}\tilde{2}\rangle - \langle \tilde{1}\tilde{2}\tilde{1}|\rho|\tilde{1}\tilde{2}\tilde{1}\rangle - \langle \tilde{2}\tilde{1}\tilde{1}|\rho|\tilde{2}\tilde{1}1\rangle\nonumber \\
&- \langle \tilde{0}\tilde{0}\tilde{2}|\rho| \tilde{0}\tilde{0}\tilde{2}\rangle - \langle \tilde{0}\tilde{2}\tilde{0}|\rho|\tilde{0}\tilde{2}\tilde{0}\rangle - \langle \tilde{2}\tilde{0}\tilde{0}|\rho|\tilde{2}\tilde{0}\tilde{0}\rangle \nonumber\\
&- \langle \tilde{1}\tilde{1}\tilde{0}|\rho| \tilde{1}\tilde{1}\tilde{0}\rangle - \langle \tilde{1}\tilde{0}\tilde{1}|\rho|\tilde{1}\tilde{0}\tilde{1}\rangle - \langle \tilde{0}\tilde{1}\tilde{1}|\rho|\tilde{0}\tilde{1}\tilde{1}\rangle \nonumber\\
&- \langle \tilde{2}\tilde{2}\tilde{1}|\rho| \tilde{2}\tilde{2}\tilde{1}\rangle - \langle \tilde{2}\tilde{1}\tilde{2}|\rho|\tilde{2}\tilde{1}\tilde{2}\rangle - \langle \tilde{1}\tilde{2}\tilde{2}|\rho|\tilde{1}\tilde{2}\tilde{2}\rangle \nonumber \\
&= \frac{2}{3} \{ \nonumber\\
& + \Re e \langle 000|\rho|111\rangle + \Re e \langle 000|\rho|222\rangle +  \Re e \langle 001|\rho|220\rangle + \Re e \langle 002|\rho|110\rangle \nonumber\\
& + \Re e \langle 002|\rho|221\rangle + \Re e \langle 010|\rho|121\rangle +  \Re e \langle 010|\rho|202\rangle + \Re e \langle 011|\rho|122\rangle  \nonumber\\
& + \Re e \langle 011|\rho|200\rangle + \Re e \langle 012|\rho|120\rangle +  \Re e \langle 012|\rho|201\rangle + \Re e \langle 020|\rho|101\rangle  \nonumber\\
& + \Re e \langle 020|\rho|212\rangle + \Re e \langle 021|\rho|102\rangle +  \Re e \langle 021|\rho|210\rangle + \Re e \langle 022|\rho|100\rangle  \nonumber\\
& + \Re e \langle 022|\rho|211\rangle + \Re e \langle 100|\rho|211\rangle +  \Re e \langle 101|\rho|212\rangle + \Re e \langle 102|\rho|210\rangle \nonumber \\
& + \Re e \langle 110|\rho|221\rangle + \Re e \langle 111|\rho|222\rangle +  \Re e \langle 112|\rho|220\rangle + \Re e \langle 120|\rho|201\rangle \nonumber \\
& + \Re e \langle 121|\rho|202\rangle + \Re e \langle 122|\rho|200\rangle - \underbrace{\sum_{i} \langle i|\rho|i\rangle}_{=1} \} \nonumber\\
\end{align}
which results in the following bound

\begin{align}
C_{GME} &\geq B_{GME}(\rho):=\frac{2}{\sqrt{3}} [ 
( \frac{3}{2}  C_{3,3} + 1 \nonumber\\
&- \sqrt{\langle 001|\rho|001\rangle\langle 220|\rho|220\rangle} - \sqrt{\langle 002|\rho|002\rangle\langle 110|\rho|110\rangle} - \sqrt{\langle 002|\rho|002\rangle\langle 221|\rho|221\rangle}\nonumber\\
&- \sqrt{\langle 010|\rho|010\rangle\langle 121|\rho|121\rangle} - \sqrt{\langle 010|\rho|010\rangle\langle 202|\rho|202\rangle} - \sqrt{\langle 011|\rho|011\rangle\langle 122|\rho|122\rangle}\nonumber\\
&- \sqrt{\langle 011|\rho|011\rangle\langle 200|\rho|200\rangle} - \sqrt{\langle 012|\rho|012\rangle\langle 120|\rho|120\rangle} - \sqrt{\langle 012|\rho|012\rangle\langle 201|\rho|201\rangle}\nonumber\\
&- \sqrt{\langle 020|\rho|020\rangle\langle 101|\rho|101\rangle} - \sqrt{\langle 020|\rho|020\rangle\langle 212|\rho|212\rangle} - \sqrt{\langle 021|\rho|021\rangle\langle 102|\rho|102\rangle}\nonumber\\
&- \sqrt{\langle 021|\rho|021\rangle\langle 210|\rho|210\rangle} - \sqrt{\langle 022|\rho|022\rangle\langle 100|\rho|100\rangle} - \sqrt{\langle 022|\rho|022\rangle\langle 211|\rho|211\rangle}\nonumber\\
&- \sqrt{\langle 100|\rho|100\rangle\langle 211|\rho|211\rangle} - \sqrt{\langle 101|\rho|101\rangle\langle 212|\rho|212\rangle} - \sqrt{\langle 102|\rho|102\rangle\langle 210|\rho|210\rangle}\nonumber\\
&- \sqrt{\langle 110|\rho|110\rangle\langle 221|\rho|221\rangle} - \sqrt{\langle 112|\rho|112\rangle\langle 220|\rho|220\rangle} - \sqrt{\langle 120|\rho|120\rangle\langle 201|\rho|201\rangle}\nonumber\\
&- \sqrt{\langle 121|\rho|121\rangle\langle 202|\rho|202\rangle} - \sqrt{\langle 122|\rho|122\rangle\langle 200|\rho|200\rangle} )\nonumber\\
&- \sqrt{\langle 001|\rho|001\rangle\langle 110|\rho|110\rangle} - \sqrt{\langle 010|\rho|010\rangle\langle 101|\rho|101\rangle}- \sqrt{\langle 100|\rho|100\rangle\langle 011|\rho|011\rangle} \nonumber\\ 
&- \sqrt{\langle 112|\rho|112\rangle\langle 221|\rho|221\rangle} - \sqrt{\langle 121|\rho|121\rangle\langle 212|\rho|212\rangle}- \sqrt{\langle 122|\rho|122\rangle\langle 211|\rho|211\rangle} \nonumber\\ 
&- \sqrt{\langle 002|\rho|002\rangle\langle 220|\rho|220\rangle} - \sqrt{\langle 020|\rho|020\rangle\langle 202|\rho|202\rangle}- \sqrt{\langle022|\rho|022\rangle\langle 200|\rho|200\rangle}  ]
\end{align}

Using this criterion with the qutrit generalization of the $GHZ$ state $|GHZ\rangle=\frac{1}{\sqrt{3}}(|000\rangle+|111\rangle+|222\rangle)$, the noise resistance is still $p_{crit}=\frac{32}{59}$ and thus also practically feasible for experiments (such as e.g. \cite{mehul}).

\section{Step-by-step calculation in three dimensions}
As an instructive manual for applying the criterion we have simulated coincidences in a $3\times3$ experiment, including a conservative estimation of the involved noise. If we label the potential outcomes in one basis by standard

\begin{align}   
\textnormal{Basis 1}: \left(
\begin{array}{ccc}
 \ket{1}_1 &  \ket{2}_1 &  \ket{3}_1
\end{array}
\right)_1\nonumber\\
\textnormal{Basis 2}: 
\left(
\begin{array}{ccc}
 \ket{1}_2 &  \ket{2}_2 &  \ket{3}_2
\end{array}
\right)_2\nonumber
\end{align}
This leads to $3x3$ correlation matrices for both bases:
\begin{align}
Corr_i=
\left(
\begin{array}{ccccccccc}
 \bra{1,1}_i\rho\ket{1,1}_i & \bra{1,2}_i\rho\ket{1,2}_i & \bra{1,3}_i\rho\ket{1,3}_i\\
 \bra{2,1}_i\rho\ket{2,1}_i & \bra{2,2}_i\rho\ket{2,2}_i & \bra{2,3}_i\rho\ket{2,3}_i\\
 \bra{3,1}_i\rho\ket{3,1}_i & \bra{3,2}_i\rho\ket{3,2}_i & \bra{3,3}_i\rho\ket{3,3}_i
\end{array}
\right).
\end{align}
Here we use only a single index $i$ as both photons are assumed to be measured in the same basis for both matrices. I.e. $\bra{1,1}_i\rho\ket{1,1}_1$ can be computed through the number of coincidence events $N_{11}$ when both parties are measuring in the first basis (simply by computing $\frac{N_{11}}{\sum_{i,j=1}^3N_{ij}}$). While this procedure will work for any quantum setup with two mutually biased bases available, let us assume for now that we perform the measurements in the position (basis 1) and momentum basis (basis 2), and using simulated data from the noise model of section (\ref{noisy}) we get the following correlation matrices:

\begin{align}
\textnormal{Corr}_1=\left(
\begin{array}{ccc}
 1015 & 23 & 9 \\
 17 & 947 & 8 \\
 9 & 28 & 1008 \\
\end{array}
\right)\label{SI_Corr1}\\ 
\textnormal{Corr}_2=\left(
\begin{array}{ccc}
 1053 & 21 & 7 \\
 29 & 1017 & 25 \\
 5 & 15 & 1023 \\
\end{array}
\right).
\end{align}
To calculate $B(\rho)$ in equation (\ref{equation2}) from the main text, which leads to the EoF, we need correlations in two unbiased bases. $C_1=\frac{\textnormal{Diag}(\textnormal{Corr}_1)}{\textnormal{Total}(\textnormal{Corr}_1)}$ is the sum of diagonal elements divided by all counts of the correlation matrix (\ref{SI_Corr1}), which gives $C_1=\frac{2970}{3064}=0.9693$. $C_2=\frac{\textnormal{Diag}(\textnormal{Corr}_2)}{\textnormal{Total}(\textnormal{Corr}_2)} = 0.9681$ gives the correlations in both bases. In the calculation of B($\rho$), we need $C_2$, which is simply the correlation in basis $2$.

The next term in B($\rho$) is subtracted from our correlations. The sum over the four variables $(m,n,l,o)$ goes each from $1$ to $3$, with the restrictions $m\neq n$,$m\neq l$,$l\neq o$,$n\neq o$, which leads to $18$ terms. For these combinations of cross-talk elements, only the first basis is used. The first term for $m=1, n=2, l=2, o=1$ gives $\sqrt{\bra{1,2}\rho\ket{1,2}\bra{2,1}\rho\ket{2,1}}=\frac{\sqrt{23\cdot17}}{N}=0.00645$, where $N=3064$ is the total number of counts of $Corr_1$ to normalize the density matrix. Similarly, the second term $m=1, n=2, l=2, o=3$ is $\sqrt{\bra{1,2}\rho\ket{1,2}\bra{2,3}\rho\ket{2,3}}=\frac{\sqrt{23\cdot8}}{N}=0.0044$; the third term $m=1, n=2, l=3, o=1$: $\sqrt{\bra{1,2}\rho\ket{1,2}\bra{3,1}\rho\ket{3,1}}=\frac{\sqrt{23\cdot9}}{N}=0.00470$ and so on. With that, we find that (with d=3)
\begin{align}
M_1=\sum_{\substack{m\neq n, m \neq l \\ l \neq o, n \neq o}} \sqrt{\langle v_m^1v_{n}^1|\rho|v_m^1v_{n}^1\rangle\langle v_{l}^1v_{o}^1|\rho|v_{l}^1v_{o}^1\rangle}=0.0852
\end{align}

The final sum has indices $i$ and $j$ each going from $1$ to $3$, with the restriction $i\neq j$. The first term with $i=1$ and $j=2$ is $\sqrt{\bra{1,2}\rho\ket{1,2}\bra{2,1}\rho\ket{2,1}}=\frac{\sqrt{55\cdot56}}{N}=0.00645$, the second term with $i=1$, $j=3$ is $\sqrt{\bra{1,3}\rho\ket{1,3}\bra{3,1}\rho\ket{3,1}}=\frac{\sqrt{9\cdot9}}{N}=0.00294$. This leads to
\begin{align}
M_2=\sum_{i \neq j}\sqrt{\langle v_i^1 v_j^1|\rho|v_i^1 v_j^1\rangle\langle v_j^1 v_i^1|\rho|v_j^1 v_i^1\rangle\,}=0.02856
\end{align}
Finally this leads to
\begin{align}
B(\rho)=\sqrt{\frac{2}{d(d-1)}} \left(d\cdot C_2 - 1 - M_1 - M_2 \right)\nonumber\\
=\sqrt{\frac{1}{3}} \left(3\cdot 0.9681 - 1 - 0.0852 - 0.02856 \right)=1.0338.                                
\end{align}
The entanglement of formation is defined in the main text in eq.~\eqref{EoF}, and if we apply our values, we find
\begin{align}
E_{oF}(\rho)\geq -\log(1-\frac{B(\rho)^2}{2})=1.1
\end{align}
That means, the state has at least $1.1$ bits of nonlocal information. This obviously implies an entanglement dimensionality of more than two (as the limit for qubits is 1). For demonstration purposes we can also find a lower bound for the entanglement dimensionality (Schmidt number) through the straightforward approach
\begin{align}
D\geq \lceil 2^{E_{oF}(\rho)} \rceil = \lceil2.4\rceil=3,
\end{align}
therefore we also show that the Schmidt number of the state is at least $3$. 

\end{document}